\documentclass{jors}

\usepackage{arxiv}
\usepackage[utf8]{inputenc} 
\usepackage[T1]{fontenc}    
\usepackage{hyperref}       
\usepackage{url}            
\usepackage{booktabs}       
\usepackage{amsfonts}       
\usepackage{nicefrac}       
\usepackage{microtype}      
\usepackage{lipsum}		
\usepackage{graphicx}
\usepackage{doi}
\usepackage[labelfont=bf]{caption}

\title{Accessing United States Bulk Patent Data with \textit{patentpy} and \textit{patentr}}

\author{James~Yu, BS \\
	School of Medicine \\
	Case Western Reserve University
	\And
	\href{https://orcid.org/0000-0002-6980-052X}{\includegraphics[scale=0.06]{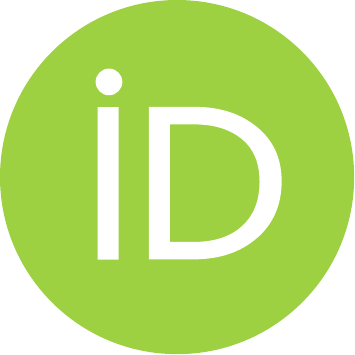}\hspace{1mm}Hayley~Beltz, MS} \\
	Department of Astronomy \\
	University of Michigan \\
	\And
	Milind Y.~Desai, MD, MBA \\
	Department of Cardiovascular Medicine \\
	Heart, Vascular \& Thoracic Institute \\
	Cleveland Clinic \\
	\And
	P\'eter~\'Erdi, PhD, DSc \\
	Department of Physics \\
	Center for Complex Systems Studies \\
	Kalamazoo College \\
	\And
    \href{https://orcid.org/0000-0003-2971-7673}{\includegraphics[scale=0.06]{orcid.pdf}\hspace{1mm}Jacob G.~Scott, MD, DPhil} \\
    Translational Hematology \& Oncology Research \\
    Lerner Research Institute \\
    Cleveland Clinic \\
    \And
    \href{https://orcid.org/0000-0003-0503-9580}{\includegraphics[scale=0.06]{orcid.pdf}\hspace{1mm}Raoul R.~Wadhwa, BA} \\
    Cleveland Clinic Lerner College of Medicine \\
    Case Western Reserve University
}

\renewcommand{\shorttitle}{\textit{patentpy} and \textit{patentr}}

\hypersetup{
pdftitle={patentpy and patentr preprint},
pdfauthor={Raoul R.~Wadhwa},
pdfkeywords={patent,USPTO,Python,R},
}

\begin{document}
\maketitle

\begin{abstract}

The United States Patent and Trademark Office (USPTO) provides publicly accessible bulk data files containing information for all patents from 1976 onward.
However, the format of these files changes over time and is memory-inefficient, which can pose issues for individual researchers.
Here, we introduce the \textit{patentpy} and \textit{patentr} packages for the Python and R programming languages.
They allow users to programmatically fetch bulk data from the USPTO website and access it locally in a cleaned, rectangular format.
Research depending on United States patent data would benefit from the use of \textit{patentpy} and \textit{patentr}.
We describe package implementation, quality control mechanisms, and present use cases highlighting simple, yet effective, applications of this software.

\end{abstract}


\section*{Introduction}

The United States Patent and Trademark Office (USPTO) hosts bulk data files for all patents published since 1976.
This data holds interesting insights into a large number of fields but is unfortunately difficult to access on large scales.
Files for each week can be hundreds of megabytes in size; additionally, files from different years are sometimes formatted distinctly.
As a result of these challenges, potential users face a high barrier to entry prior to taking advantage of the available data.
The \textit{patentpy} and \textit{patentr} packages simplify accessing USPTO data by providing programmatic interfaces that return it in a single rectangular, tidy~\cite{tidy-data} format and significantly reduce its storage size.

\indent The \textit{patentpy} package provides a Python interface with the functionality implemented using a combination of Python and C++.
The \textit{patentr} package does the same with an R interface.
Since the two packages share functionality, they also share a C++ code base, with each depending on existing XML libraries in the respective language.
Of note, Python and R both boast portability across multiple operating systems, making the corresponding packages easily available to a large number of users.

\indent Currently, the most authoritative reference for patent analytics is the World Intellectual Property Organization (WIPO) Manual.~\cite{wipo-manual}
According to the WIPO Manual, multiple databases host patent data for exploration.
These include \href{https://www.lens.org/lens/}{The Lens}, \href{https://patentscope.wipo.int/search/en/search.jsf}{Patentscope}, \href{http://worldwide.espacenet.com/?locale=en_EP}{espacenet}, \href{http://lp.espacenet.com/}{LATIPAT}, \href{http://www.epo.org/searching/free/ops.html}{EPO Open Patent Services}, and \href{http://www.google.com/patents}{Google Patents}, among others.
Most of these web services provide user-friendly, graphical point-and-click interfaces.
While initially easier to use, the time and effort scales proportionately to the number and complexity of searches conducted.
Thus, these services become cumbersome when attempting to access large amount of patent data spanning over many years.
In contrast, a programmatic interface would require a higher initial investment, but iteration would save significant additional effort for additional searches.
This is the specific niche that \textit{patentpy} and \textit{patentr} address.

\indent To demonstrate the utility of these packages, we will collect an arbitrarily selected set of patent data and conduct a preliminary analysis to answer four research questions: (1) how many patents are issued weekly; (2) which IPC classes grow fastest; (3) which IPC classes take longest to issue after application submission; and (4) whether the time between application and issuance changes over time.
We hope these examples highlight that these packages can be effectively combined with available software packages to rapidly answer questions of interest.
For the arbitrary dataset, we will pretend that the first 8 weeks of a year have special significance, and that we would like to focus our attention on these weeks for the first 5 years of available patent data.
Collecting this data takes a single line of code for both packages using \texttt{get\_bulk\_patent\_data}.

\begin{figure}[b!]
    \centering
    \includegraphics[width=\textwidth]{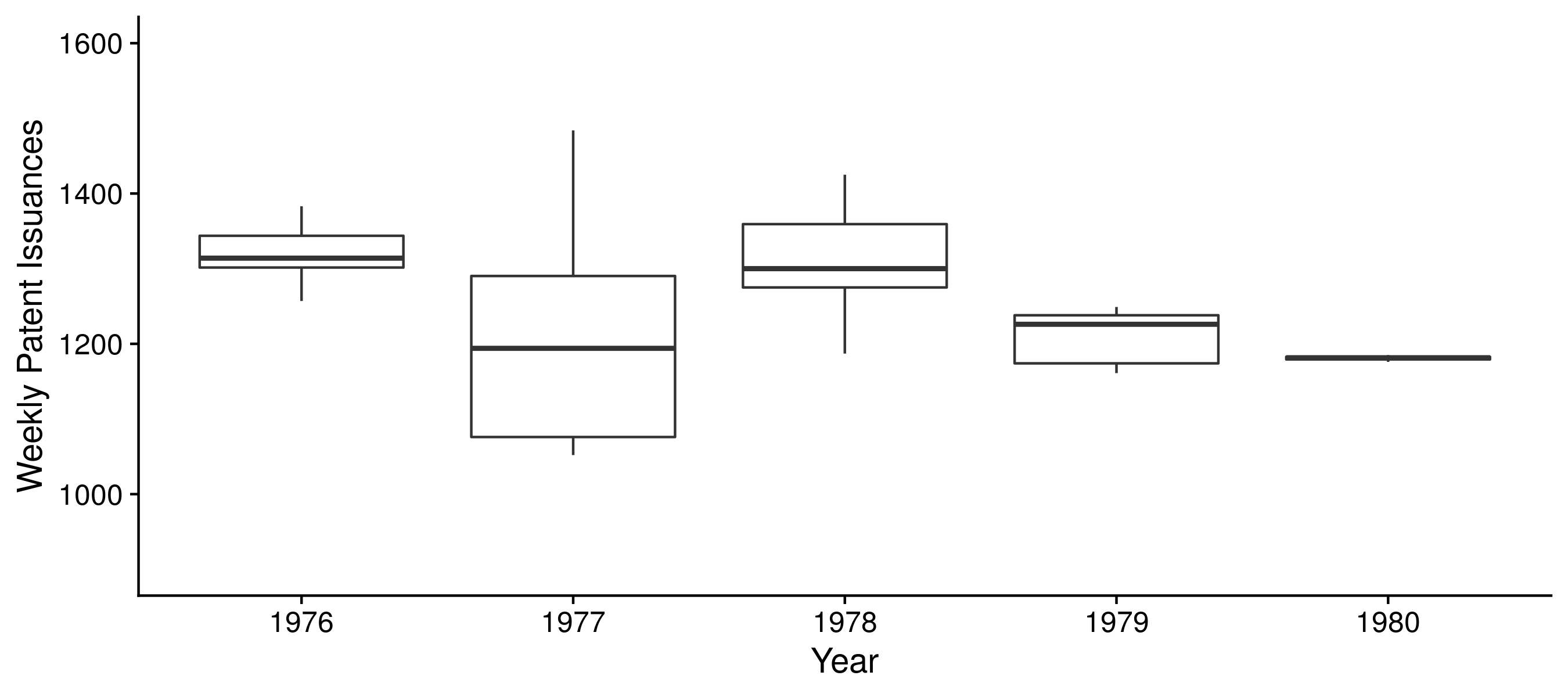}
    \caption{Distributions of number of patents issued weekly for the first 8 weeks between 1976 and 1980, inclusive.}
    \label{fig:weekly-box}
\end{figure}

\indent First, let us ask how many patents were issued weekly in the first 8 weeks of each year.
We process the data by extracting the year from each patent's issue date, aggregating and counting the number of patents issued each week, and feeding the resulting data to a visualization library.
A boxplot of the weekly aggregated patent issue counts split by year results in Figure~\ref{fig:weekly-box}.
We see that roughly 1100-1400 patents were consistently granted in the first 8 weeks of the year between 1976 and 1980, inclusive.

\begin{figure}[t!]
    \centering
    \includegraphics[width=\textwidth]{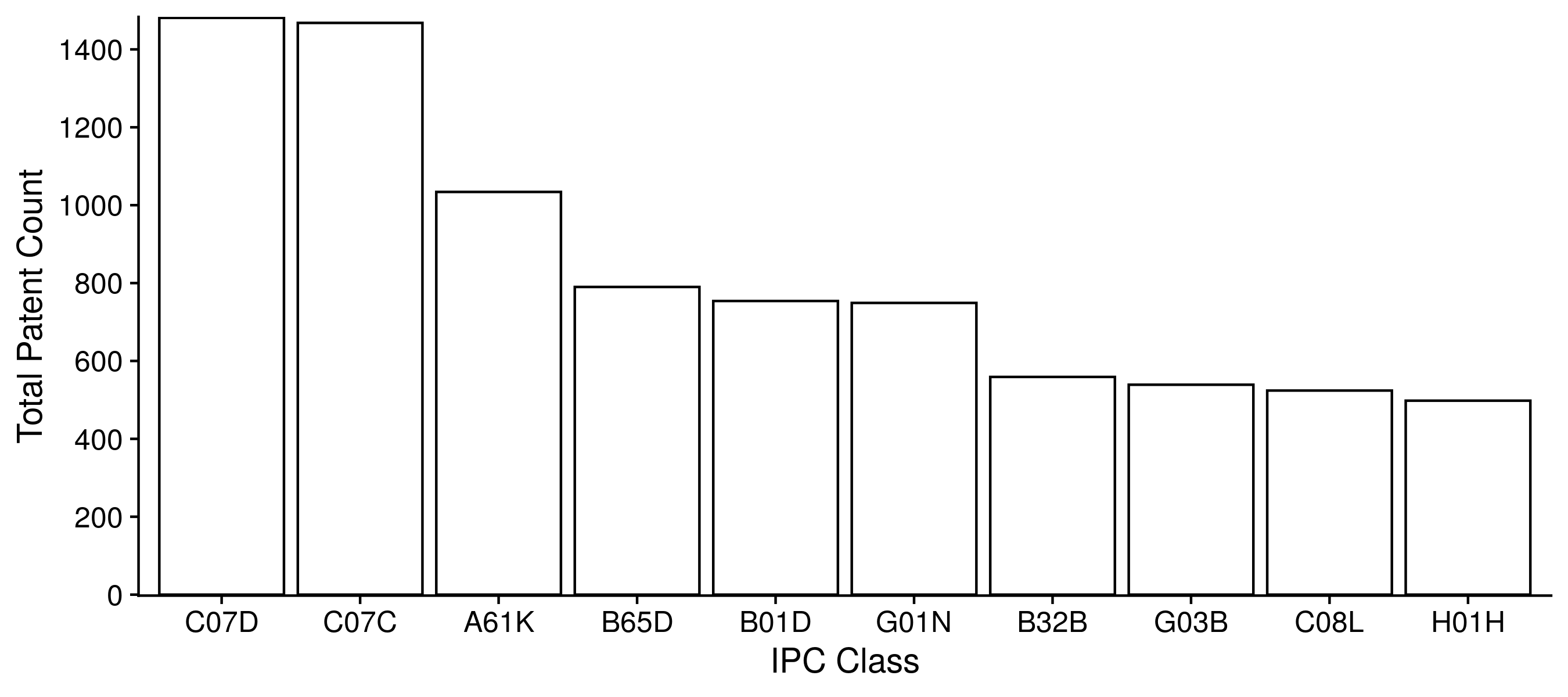}
    \caption{Number of patents issued in each International Patent Classification (IPC) class. Only counts for the 10 most frequent classes are visualized. Only patents issued in the first 8 weeks of the years between 1976 and 1980, inclusive, are included.}
    \label{fig:class-count}
\end{figure}

\indent Second, let us ask which classes of patents are growing the fastest.
The USPTO includes classifications based on the hierarchical International Patent Classification (IPC) system.
The USPTO reports the International Patent Classification (IPC); this WIPO system hierarchically classifies patents according to their relevant scientific discipline and is updated annually.~\cite{wipo-ipc}
From our dataset of patents from the first 8 weeks of each of 5 years, we count the most commonly appearing classes and visualize the 10 most frequent results in Figure~\ref{fig:class-count}.
The \texttt{C} class corresponds to ``Chemistry; Metallurgy'', within which \texttt{C07} corresponds to ``Organic Chemistry'', within which \texttt{C07D} corresponds to ``Heterocyclic Compounds''; \texttt{C07C} corresponds to ``Acyclic or Carbocyclic Compounds''.
Thus, the two fastest growing IPC classes in our dataset were both related to the field of organic chemistry.

\begin{figure}[t!]
    \centering
    \includegraphics[width=\textwidth]{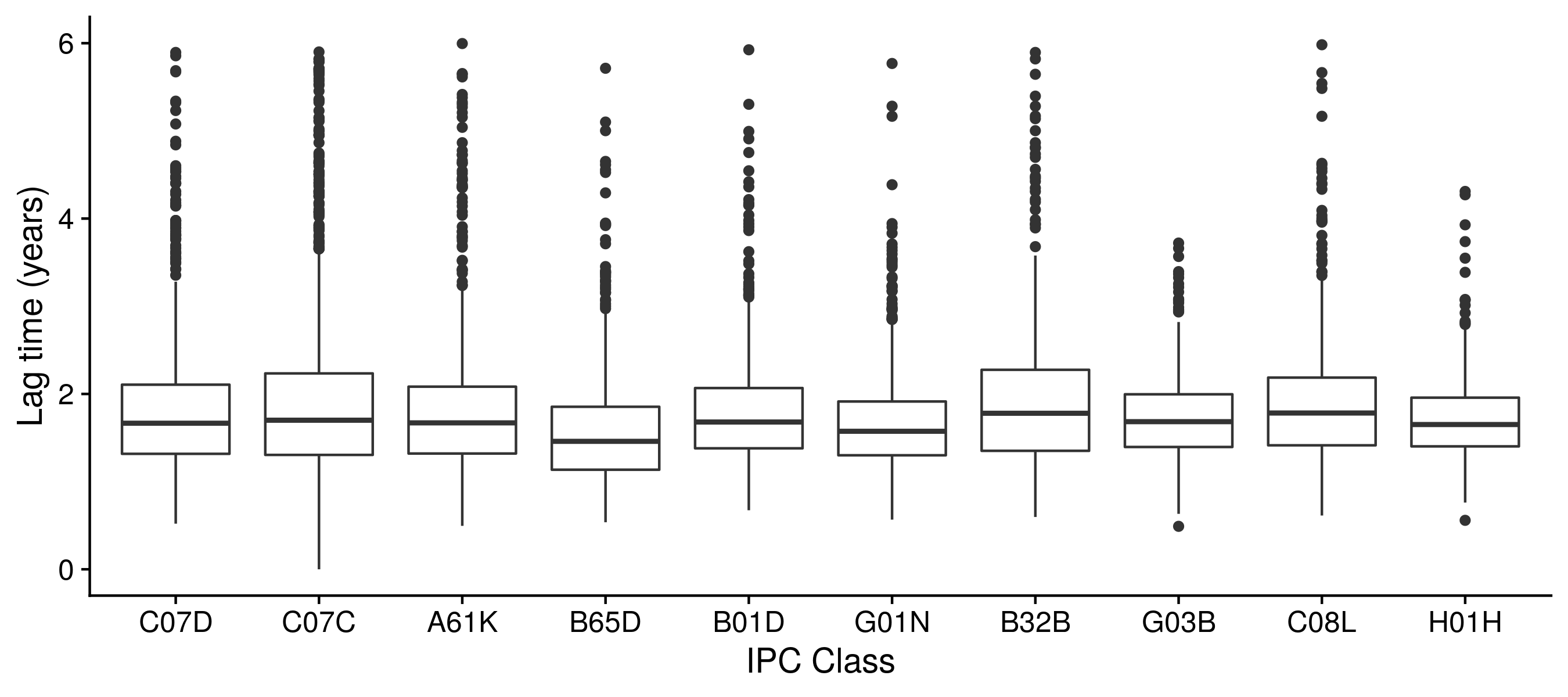}
    \caption{Distributions for time between patent application date and issue date, grouped by IPC class. Only patents issued in the first 8 weeks of the years between 1976 and 1980, inclusive, were included.}
    \label{fig:lag-class}
\end{figure}

\indent Third, we can continue our exploration of the IPC classes by asking whether patents in different classes undergo reviews of different durations prior to issuance.
After a patent application is submitted, it must first be manually approved and classified before being granted.
We denote this time between submission and issuance as the lag time.
For the sake of simplicity, we filter out patents that do not belong to one of the 10 classes shown in Figure~\ref{fig:class-count}.
We calculate the lag time between application and issuance, then visualize the results as boxplots analogous to Figure~\ref{fig:weekly-box}.
These results are shown in Figure~\ref{fig:lag-class}.
The lag time distribution for each of the 10 IPC classes is skewed right with a median duration under 2 years.

\begin{figure}[t!]
    \centering
    \includegraphics[width=\textwidth]{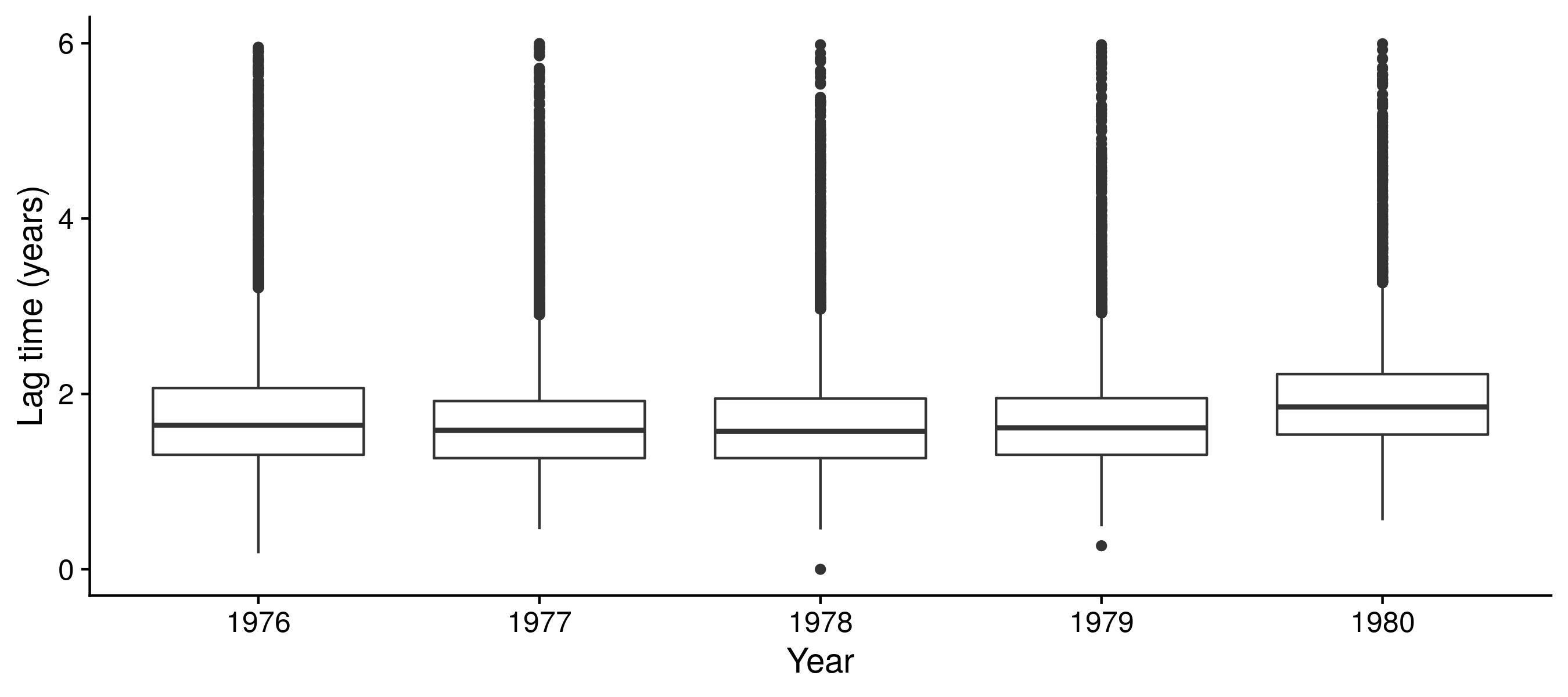}
    \caption{Distributions for time between patent application date and issue date, grouped by issue year. Only patents issued in the first 8 weeks of the years between 1976 and 1980, inclusive, were included.}
    \label{fig:lag-year}
\end{figure}

\indent Fourth, we can ask whether the lag time distribution changes across years.
To answer this question, we calculate the lag time between application and issuance - this time without filtering out any patents from our collected database - and visualize the results as boxplots (Figure~\ref{fig:lag-year}).
Here again, we note that the distributions are skewed right with medians under 2 years.
However, we also note that the boxplot for 1980 is slightly shifted upward.
Following this forward may reveal a trend of increasing lag time as years progress.
Pursuing this research question remains an open question for interested readers.


\section*{Implementation and architecture}

The \textit{patentpy} package was implemented in Python 3 and uses \textit{pybind11} to incorporate C++.~\cite{pybind}
The C++ code parses USPTO files with the TXT extension and the \textit{lxml} library parses USPTO XML files.~\cite{lxml}
The data frame structure provided by the \textit{pandas} library encapsulates rectangular data returned to the user.~\cite{pandas}
The \textit{patentr} package was implemented in R 4 and uses Rcpp to incorporate C++.~\cite{rcpp}
The C++ code parses USPTO files with the TXT extension and the \textit{xml2} libary parses USPTO XML files.~\cite{xml2}
Tibbles form the core data structure for \textit{patentr} to store and return rectangular data to the user.
Tibble functionality depends on the \textit{tidyverse} set of R packages.~\cite{tidyverse}

\indent Both packages extract the same data for each patent and format it the same way.
Specifically, both return a rectangular data object with columns representing a unique patent identifier (WKU), title, application date, issue date, inventor(s), assignee(s), IPC class(es), reference(s), and claim(s).
Additionally, multiple values in the same column (e.g. multiple references), are delimited by semi-colons in both programs, except for claims about the patent's novelty made by the preparer, for which text formatting is preserved.
Thus, when returned data from each package is saved as a comma-separated values (CSV) file, outputs from both packages should be roughly equivalent.

\indent The main entry point of the programmatic interface for each package is the \texttt{get\_bulk\_patent\_data} function, which downloads, parses, formats, and returns USPTO bulk patent data based on the USPTO Green, Yellow, and Red books.~\cite{uspto-books}
Due to the portability of Python and R, both packages are available across multiple operating systems and architectures.~\cite{pylang,rlang}


\section*{Quality control}

Automated unit testing is implemented for \textit{patentpy}, supported by Travis CI continuous integration.
Package functionality has been checked on the Windows, Mac, Ubuntu, and Debian operating systems.
Runtime errors are caught and returned with appropriate error messages based on error type.
Codecov implements code coverage.
Users can manually confirm package functionality by locally running sample code provided in the README file or by downloading and locally running test scripts, both of which are provided in \textit{patentpy}'s code repository.

\indent Automated unit testing is implemented for \textit{patentr}, supported by Travis CI and AppVeyor CI continuous integration.
Package functionality has been checked on the Windows, Mac, Ubuntu, and Debian operating systems with regular checking also completed by the Comprehensive R Archive Network (CRAN).
Codecov implements code coverage.
The package is automatically checked with unit tests upon installation from CRAN.
Users can manually confirm package functionality by locally running sample code provided in package documentation and the repository README file or by locally running unit tests using the testthat package.~\cite{testthat}


\section*{Availability and dependencies}

\textit{patentpy} is available in the Python programming language (version $\geq3.5$) via the Python Package Index (PyPI) and \textit{patentr} is available in the R programming language (version $\geq4.0$ via the Comprehensive R Archive Network (CRAN); the source code of each package is publicly available at \url{https://github.com/JYProjs/patentpy} and \url{https://github.com/JYProjs/patentr}, respectively.
\textit{patentpy} depends on \textit{pandas} (version $\geq1.2$) and \textit{lxml} (version $\geq4.6$).~\cite{lxml,pandas}
\textit{patentr} depends on \textit{covr} (version $\geq3.5.1$), \textit{dplyr} (version $\geq1.0.2$), \textit{knitr} (version $\geq1.33$), \textit{lubridate} (version $\geq1.7.9$), \textit{magrittr} (version $\geq2.0$), \textit{progress} (version $\geq1.2.2$), \textit{Rcpp} (version $\geq1.0.5$), \textit{readr} (version $\geq1.4.0$), \textit{rlang} (version $\geq0.4.7$), \textit{rmarkdown} (version $\geq2.9$), \textit{testthat} (version $\geq3.0.4$), \textit{utils} (version $\geq4.0$), and \textit{xml2} (version $\geq1.3.2$).~\cite{covr,tidyverse,knitr,magrittr,progress,rcpp,rlang-pkg,rmarkdown,xml2}

\section*{Reuse potential}

Given the large amount of rich data available for patents and the significant financial value that patents can hold, multiple research avenues have already been pursued with tools analogous to \textit{patentpy} and \textit{patentr}.
However, with their introduction to the research literature, each individual project should no longer require writing an entire base of code, saving time and resources.
The explored avenues of research range from studying international patent families~\cite{patent-family} and the network structure of patent citations~\cite{patent-network} of to patent protection for new forms of technology~\cite{patent-protect1,patent-protect2,patent-protect3,patent-protect4} and management of patent portfolios.~\cite{patent-portfolio1,patent-portfolio2}
Although USPTO bulk data files are identified by patent publication date, the \textit{patentpy} and \textit{patentr} packages could be used for data collection in any of the aforementioned research avenues.

\indent Within the field of scientometrics, further analysis of co-citation networks would be useful in the sub-field of bibliometrics.
Studying innovation within specific scientific fields over time and identifying crucial patents to the development of new technologies would also be useful.
In particular, identifying prominent patent suites and oligopolies of corporate patent assignees within scientific fields could result in illuminating research at the intersection of scientometrics and economics.
Outside of academia, scientists in industry could use patent data to study innovation and guide research development efforts based on evidence-based predictions.
The potential aforementioned pathways would all benefit from the use of either \textit{patentpy} or \textit{patentr} for data collection purposes.

\indent The presented packages are well complemented by available software for network analysis, topological data analysis, and machine learning.
As previously mentioned, patent data can be structured as a citation network or a co-citation network.
These structures can then be analyzed with the existing foundation of network and graph theory to gain insight into the evolution \textit{of} the patent network over time and the change in dynamic metrics \textit{on} the patent network.
The igraph package implements this functionality and would work well with \textit{patentpy} and \textit{patentr} in a single pipeline.~\cite{igraph}
Topological data analysis is a growing field, with an increasing number of applications of the Mapper algorithm~\cite{mapper} and persistent homology~\cite{phom} appearing in the scientific literature.
Analyzing the topology of patent networks could be completed well with packages like scikit-tda in Python and TDAstats in R.~\cite{scikit-tda,TDAstats}

\indent Machine learning pipelines have the ability to incorporate network analytics and topological features; even without these, machine learning has high potential use with patent data.
Supervised learning and natural language processing can be combined to automate classification of patents based on title, references, and claim text.
Reinforcement learning can be combined with ranking algorithms to measure the importance of a patent over time.~\cite{reinforce-patent1,reinforce-patent2}
Even unsupervised learning can be applied to find patent clusters across IPC classes and identify potential relationships between otherwise unrelated patents.
The scikit-learn package in Python and the caret and tidymodels packages in R would be candidate packages to perform such analyses on data acquired via the packages introduced in this report.~\cite{scikit-learn,caret,tidymodels}
Useful extensions and improvements to \textit{patentpy} and \textit{patentr} would include the addition of a shared interface to coordinate pipelines across the aforementioned packages, depending on the technique used to answer a research question.

\indent Contributions to \textit{patentpy} and \textit{patentr} are welcome and can be made via the code repository for each package.
Contributors can open pull requests to incorporate additions or changes.
Issues can also be opened to make feature requests and identify bugs.
The authors of this report bear responsibility for maintaining \textit{patentpy} and \textit{patentr}, and we welcome community involvement.
Support mechanisms include continuous integration with the Travis and AppVeyor tools to validate package function over time, particularly as the programming languages and software dependencies evolve.


\section*{Acknowledgements}

This project was supported by the NIH R37 grant CA244613.
The authors declare that they have no competing interests.



\end{document}